\documentclass[aps,prl,twocolumn,showpacs,superscriptaddress,floatfix]{revtex4-1}
\usepackage{bm}
\usepackage{amssymb}
\usepackage{graphicx}
\usepackage{amsmath}
\usepackage{dcolumn}
\usepackage[pdftex]{epsfig}
\usepackage{color}
\usepackage[german,english]{babel}
\usepackage{psfrag}
\usepackage{hyperref}
\usepackage{subfigure}
\usepackage{color}

\newcommand{\beq}{\begin{eqnarray}}
\newcommand{\eeq}{\end{eqnarray}}
\newcommand{\beqa}{\begin{equation}}
\newcommand{\eeqa}{\end{equation}}
\newcommand{\msu}{\uparrow}			
\newcommand{\msd}{\downarrow}			
\DeclareMathOperator{\Tr}{Tr}
\newcommand{\expec}[1]{\langle#1\rangle}

\newcommand{\ms}[1]{\textcolor{black}{#1}}

\newcommand{\eV}{\ \mathrm{eV}}

\begin{document}

\title{Optimal Hubbard models for materials with nonlocal Coulomb interactions: graphene, silicene and benzene}
\author{M. Sch\"uler}
\email{mschueler@itp.uni-bremen.de}
\author{M. R\"osner}
\author{T. O. Wehling}
\affiliation{Institut f{\"u}r Theoretische Physik, Universit{\"a}t Bremen, Otto-Hahn-Allee 1, 28359 Bremen, Germany}
\affiliation{Bremen Center for Computational Materials Science, Universit{\"a}t Bremen, Am Fallturm 1a, 28359 Bremen, Germany}
\author{A. I. Lichtenstein}
\affiliation{Institut f{\"u}r Theoretische Physik,
Universit{\"a}t Hamburg, Jungiusstra{\ss}e 9, D-20355 Hamburg,
Germany}
\author{M. I. Katsnelson}
\affiliation{Radboud University of Nijmegen, Institute for
Molecules and Materials, Heijendaalseweg 135, 6525 AJ Nijmegen,
The Netherlands}

\date{\today}

\begin{abstract}
To understand how nonlocal Coulomb interactions affect the phase diagram of correlated electron materials, we report on a method to approximate a correlated lattice model with nonlocal interactions by an effective Hubbard model with on-site interactions $U^*$ only. The effective model is defined by the Peierls-Feynman-Bogoliubov variational principle. We find that the local part of the interaction $U$ is reduced according to $U^*=U-\bar V$, where $\bar V$ is a weighted average of nonlocal interactions. For graphene, silicene and benzene we show that the nonlocal Coulomb interaction can decrease the effective local interaction by more than a factor of 2 in a wide doping range.
\end{abstract}



\maketitle
Low dimensional $sp$-electron systems like graphene \cite{wehling_strength_2011,Castro-Neto_RMP12,Misha_book}, systems of adatoms on semiconductor surfaces, such
as Si(111):X with X=C, Si, Sn, Pb \cite{Biermann_2013}, Bechgaard salts or aromatic molecules \cite{pariser_semiempirical_1953, Pople_1955} and polymers \cite{elec_band_1999,soos_band_1993} feature simultaneously strong local and nonlocal Coulomb interactions. In graphene for instance, the on-site interactions $U/t\sim 3.3$, the nearest neighbor Coulomb repulsion $V/t\sim 2$ as well as further sizable nonlocal Coulomb terms exceed the nearest neighbor hopping $t=2.8\eV$ \cite{wehling_strength_2011}. Considering on-site interactions $U/t\sim 3.3$ alone would put graphene close to the boundary of a gapped spin-liquid \cite{Muramatsu_Nature10}, which could be even crossed by applying strain on the order of a few percent \cite{wehling_strength_2011}. It is currently unclear, whether \cite{MacDonald_PRB2011} or not \cite{honerkamp_CDW_SDW_PRL08,scherer_interacting_2012} nonlocal Coulomb interaction stabilize the semimetallic Dirac phase in graphene. To rephrase the problem: It is unclear which Hubbard model with strictly local interactions would yield the best approximation to the ground state of graphene. To judge the stability of the Dirac electron phase in graphene but also to understand Mott transitions on surfaces like Si:X (111), a quantitative well defined link from models with local and nonlocal Coulomb interactions to those with purely local interactions is desirable.

In this letter, we present a method to map a generalized Hubbard model with nonlocal Coulomb interactions onto an effective Hubbard model with on-site interactions $U^*$ only. For graphene, silicene and benzene we show \ms{that} nonlocal terms reduce the effective on-site interaction by more than a factor of two in a wide doping range around half filling. Thus, nonlocal Coulomb interactions are found to stabilize the Dirac electron phases in graphene and silicene against spin-liquid and antiferromagnetic phases. In the almost empty and nearly filled case we find, however, that even strictly repulsive nonlocal Coulomb interactions can effectively increase the local interactions.

The starting point is the extended Hubbard model
\begin{align}
H = -\sum_{i,j,\sigma} t_{ij} c_{i\sigma}^\dagger c_{j\sigma} + U\sum_i n_{i\msu}n_{i\msd} + \frac{1}{2} \sum_{\stackrel{i\neq j}{\sigma,\sigma'}}V_{ij} n_{i\sigma}n_{j\sigma'}, \label{eq:origHam}
\end{align}
where $t_{ij}$ are the hopping matrix elements. $U$ and $V_{ij}$ are the local and nonlocal Coulomb matrix elements, respectively. The goal is to map the Hamiltonian (\ref{eq:origHam}) onto the effective model
\begin{align}
H^* = -\sum_{i,j,\sigma} t_{ij} c_{i\sigma}^\dagger c_{j\sigma} + U^*\sum_i n_{i\msu}n_{i\msd}.
\end{align}
The effective on-site interaction $U^*$ shall be chosen such that the canonical density operator $\rho^*=1/Z^* e^{-\beta H^*}$ of the auxiliary system, where $Z^*=\Tr \left\{ e^{-\beta H^*} \right\}$ is the partition function, approximates the exact density operator $\rho$ derived from $H$ as close as possible. This requirement leads to the Peierls-Feynman-Bogoliubov variational principle \cite{Peierls_1938, Bogoliubov_1958, Feynman_1972} for the functional
\begin{align}
\tilde{\Phi}[\rho^*] = \Phi^* + \expec{H-H^*}^*,
\label{eq:Ustar_functional}
\end{align}
where $\Phi^* = -\frac{1}{\beta} \ln Z^*$ is the free energy of the auxiliary system. $\expec{\dots}^*=\Tr \rho^* (\cdots)$ denotes thermodynamic expectation values with respect to the auxiliary system. In the case of $\rho^*=\rho$ the functional $\tilde{\Phi}[\rho^*]$ becomes minimal and coincides with the free energy. The optimal $U^*$ is thus obtained for minimal $\tilde{\Phi}[\rho^*]=\tilde{\Phi}[U^*]$:
\begin{align}
\partial_{U^*}\tilde{\Phi}[U^*] =0.
\label{eq:extremal_principle}
\end{align}
By evaluating Eq. (\ref{eq:extremal_principle}) one finds
\begin{align}
U^*=U+ \frac{1}{2} \sum_{\stackrel{i\neq j}{\sigma,\sigma'}}V_{ij}  \frac{\partial_{U^*}\expec{n_{i\sigma}n_{j\sigma'}}^*}{\sum_l  \partial_{U^*} \expec{n_{l\msu}n_{l\msd}}^*}. \label{eq:Ustar}
\end{align}

\begin{figure}
\begin{center}
\mbox{
\includegraphics[width=1\columnwidth]{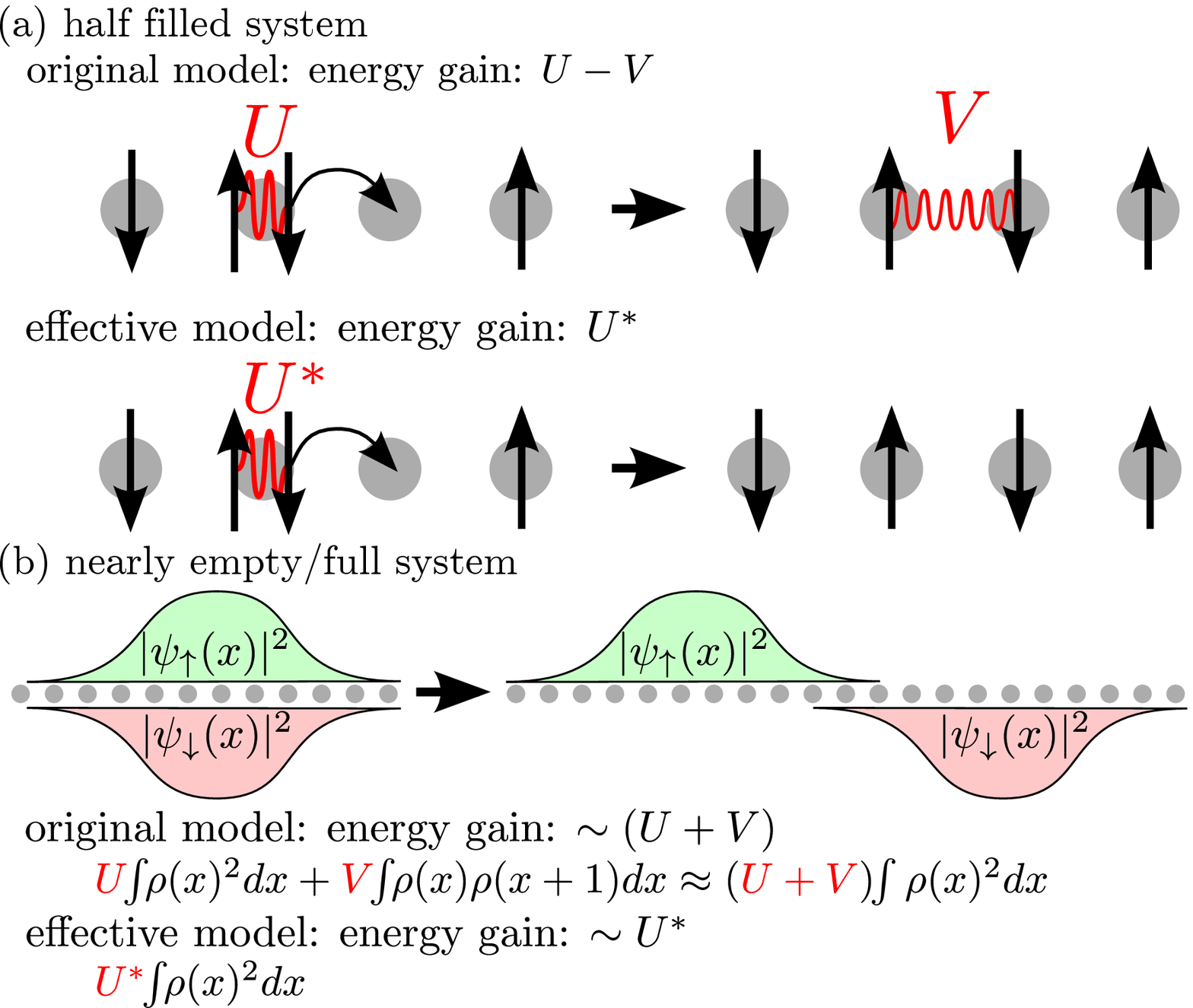}
}
\end{center}
\caption{(Color online) Illustration of the physical process underlying Eq. (\ref{eq:Ustar}). (a) Half filled system: An electron hops from a doubly occupied site to an empty one, gaining an energy $(U-V)$ in the original model and $U^*$ in the effective model.  (b) Nearly empty/full system: Wave packets of spin up and down electrons/holes ($\rho(x)=|\psi_{\msu,\msd}(x)|^2$) are separated to the farthest possible position. If the packets are much wider than the lattice spacing, the energy gained in the original model is $\sim (U+V)$ and $\sim U^*$ in the effective model.}
\label{fig:physical}
\end{figure}

This rule presents a central result of this letter and has an intuitive physical interpretation (\mbox{Fig. \ref{fig:physical}}): Increasing the on-site term $U^*$ reduces the double occupancy $\expec{n_{i\msu}n_{i\msd}}^*$ and pushes away electrons approaching an already occupied site $i=0$ to neighboring sites. In case of purely local Coulomb interactions there is a Coulomb energy gain of $U^*$ upon suppressing the double occupancy. When there are, however, nonlocal Coulomb interactions with surrounding lattice sites $j$, the displaced electrons raise the energy of the system by terms proportional to $V_{0j}$. For a system at half filling with one doubly occupied site this process is illustrated in Fig. \ref{fig:physical}a). In this case, it is obvious that the Coulomb energy gain due to the electron displacement in the original and the auxiliary model become energetically equivalent for $U^*=U-V$. We will show that this picture applies well for graphene, silicene and benzene in a wide doping range.

For a translationally invariant system, the local part of the interaction $U$ is reduced according to $U^*=U-\bar V$, where 
\begin{align}
\bar V=-\sum_{\stackrel{j\neq 0}{\sigma'}}V_{0j}  \frac{\partial_{U^*}\expec{n_{0\msu}n_{j\sigma'}}^*}{\partial_{U^*} \expec{n_{0\msu}n_{0\msd}}^*}. \label{eq:weight}
\end{align}
The conservation of the total electron number $N$ leads to the sum rules ${\sum_{j\sigma} \expec{n_{0\msu} n_{j\sigma}}^* = N/2}$ and ${\partial_{U^*} \expec{n_{0\msu}n_{0\msd}}^*=-\sum_{j\neq 0,\sigma} \partial_{U^*}\expec{n_{0\msu} n_{j\sigma}}^*}$. Thus,  $\bar V$ is a weighted average of the nonlocal Coulomb interactions. Under the assumption that an increasing $U^*$ displaces electrons only to next neighbors, we find
$\partial_{U^*} \expec{n_{0\msu} n_{0\msd}}^* = -N_{n} \partial_{U^*} \sum_\sigma \expec{n_{0\msu}n_{1 \sigma'}}^*$,
where $N_n$ is the coordination number. Eq. (\ref{eq:Ustar}) then yields 
\begin{align}
U^* = U - V_{01}. \label{eq:approx}
\end{align} 
This gives an estimate for the effective Coulomb interaction, without the need of numerical calculations but it follows from a severe approximation. The following numerical calculations show, however, that in a wide doping range around half filling Eq. (\ref{eq:approx}) leads to values close to the exact ones (shown in Table \ref{tab:results}). Then, the nonlocal Coulomb interaction reduces the effective on-site interaction and therefore stabilizes the Fermi sea against transitions e.g. to a Mott insulator. Nevertheless, situations with negative $\bar V$ can be constructed, as will be demonstrated for systems with nearly empty or almost filled bands further below.  

When the approximation that electrons are only displaced to next neighbors is dropped, the derivatives of the correlation functions have to be calculated explicitly. This can be done approximately within the dynamical mean field theory \cite{georges_dynamical_1996} and diagrammatic extensions like the Dual-Fermion approach \cite{rubtsov_dual_2008}. In certain cases also numerically exact calculations of the nonlocal charge correlation functions for instance by means of exact diagonalization (ED), determinant quantum Monte Carlo (DQMC \cite{blankenbecler_monte_1981}) or density-matrix renormalization group methods (e.g. \cite{noack_correlations_1994}) are possible.

In the following, we consider graphene, silicene and benzene by means of DQMC and ED. We used the DQMC implementation ``QUantum Electron Simulation Toolbox'' (\textsc{quest} 1.3.0 \footnote{A. Tomas, C-C. Chang, Z-J. Bai, and R. Scalettar, \textsc{quest} code (\url{http://quest.ucdavis.edu/})}) on a super cell to obtain the charge correlation functions that enter Eq. (\ref{eq:Ustar}) for graphene and silicene at half filling. Furthermore, a different DQMC implementation \footnote{private communication with Fakher F. Assaad (University of W{\"u}rzburg) \cite{Assaad08_rev}} was used to verify the results of the \textsc{quest} package. The Hubbard model with less than 8-9 sites can also be solved by exact diagonalization (ED). In this case a comparison with data obtained with DQMC shows excellent agreement \cite{suppl}.

To calculate $U^*$ for realistic systems, we introduce values for the Coulomb interactions in the original model defined by Eq. (\ref{eq:origHam}). For graphene and silicene these values are calculated with the constrained random phase approximation (cRPA) \cite{aryasetiawan_frequency-dependent_2004} like in \cite{wehling_strength_2011}. For benzene, we use values from \cite{bursill_optimal_1998}, which are obtained by fitting $U$ and $t$ to experimental spectra and calculating $V_{ij}$ by Ohno interpolation \cite{ohno_remarks_1964}, which reads
\begin{align}
V_{ij}(\varepsilon) = \frac{U}{\sqrt{1+ (\alpha \varepsilon r_{ij})^2}} \label{eq:ohno}
\end{align}
with $\alpha=U/e^2$. The nonlocal Coulomb interaction can be tuned by an additional variable screening $\varepsilon$ ranging from $0$ to $\infty$. $\varepsilon=\infty$ corresponds to purely local interactions and $\varepsilon=0$ to ultimately nonlocal interactions with matrix elements not decaying with distance between sites. $\varepsilon=1$ corresponds to the model of benzene proposed in \cite{bursill_optimal_1998}.
For all systems the values of the initial Coulomb interactions are given in Table \ref{tab:results}.

\begin{figure}
\begin{center}
\includegraphics[width=1\columnwidth]{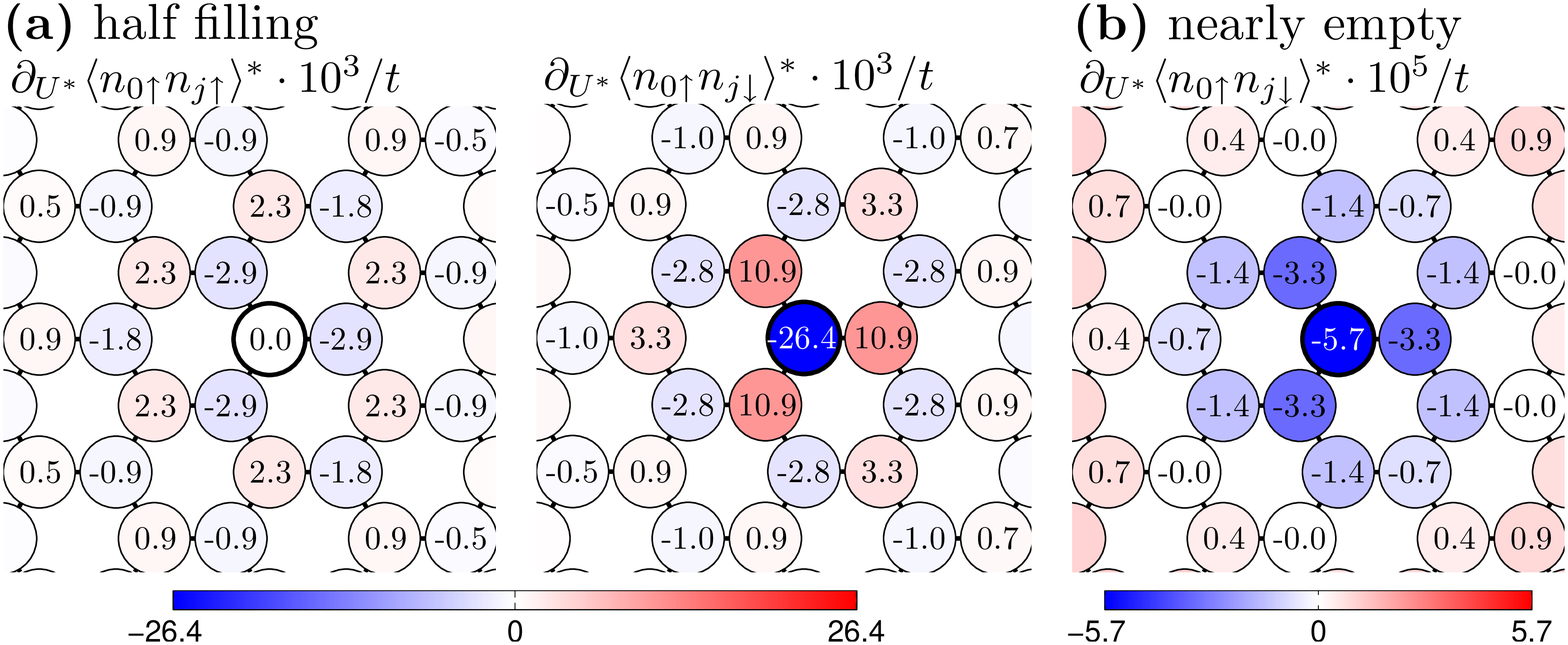}
\end{center}
\caption{(Color online) (a) Derivatives of the correlation functions $\expec{n_{0\msu} n_{j\sigma}}$ with respect to $U^*$ at $U^*=2t$ for half filled graphene (16x16 unit cells). Each circle corresponds to one carbon atom. The thick drawn circles indicate the lattice site with index $i=0$. (b) Correlation function $\partial_{U^*}\expec{n_{0\msu}n_{j\msd}}^*$ for nearly empty honeycomb lattice (2 electrons in $5\times5$ super cell). In addition, we find $\partial_{U^*}\expec{n_{0\msu}n_{j\msu}}^*=0$ as it must be for a singlet ground state (not shown here). }
\label{fig:corrFunc}
\end{figure}

For a honeycomb lattice at half filling the $U^*$ derivatives of the correlation functions $\partial_{U^*}\expec{n_{0\msu}n_{j\sigma'}}^*$ at $U^* = 2t$ are shown in Fig. \ref{fig:corrFunc}a) \footnote{DQMC calculations with 16x16 unit cells, $\beta=9t$, $\Delta\tau=0.05$ and 3000 measurement sweeps. To overcome the statistical noise, we fit the correlation functions with polynomials of rank 4 in $U^*$ and evaluate the derivative analytically.}. In this particular case, $\partial_{U^*}\expec{n_{0\msu}n_{j\sigma'}}^*$ changes sign with both sublattice and spin indices. Generally, $|\partial_{U^*}\expec{n_{0\msu}n_{j\msd}}^*|$ with opposite spins exceeds the equal spin case $|\partial_{U^*}\expec{n_{0\msu}n_{j\msu}}^*|$. The derivatives decrease clearly with the distance between the sites $i=0$ (thick drawn circles in the middle) and $j$. Thus, our numerical calculations show that upon increasing $U^*$ double occupancy is indeed reduced by displacing electrons to close by neighboring sites and support the scenario suggested in Fig. \ref{fig:physical}a).

The resulting values of the effective local Coulomb interaction $U^*$ for graphene, silicene and benzene are summarized in Table \ref{tab:results}. The local Coulomb interaction is decreased by a factor of larger than two in all cases. For both, graphene and silicene the renormalized on-site interactions are far away from the transition to a gapped spin liquid at $U^*/t=3.5$ \cite{Muramatsu_Nature10}. The Dirac semimetal phase is thus stabilized by the nonlocal Coulomb interactions. We obtain the strongest renormalization of the on-site interaction for benzene. This is mostly due to the different ratio between local and nonlocal Coulomb interactions in benzene, $V_{01}/U=0.72$, as compared to $V_{01}/U=0.56$ for graphene \footnote{A comparison with a graphene result for 8x8 unit cells, which yields the same ratio $U/U^*$, rules out finite size effects of the DQMC calculations.} or $V_{01}/U=0.55$ for silicene. 

\begin{figure}[t]
\begin{center}
\mbox{
\psfrag{inft}{$\infty$}
\psfrag{A}{$\varepsilon$}
\includegraphics[width=1\columnwidth]{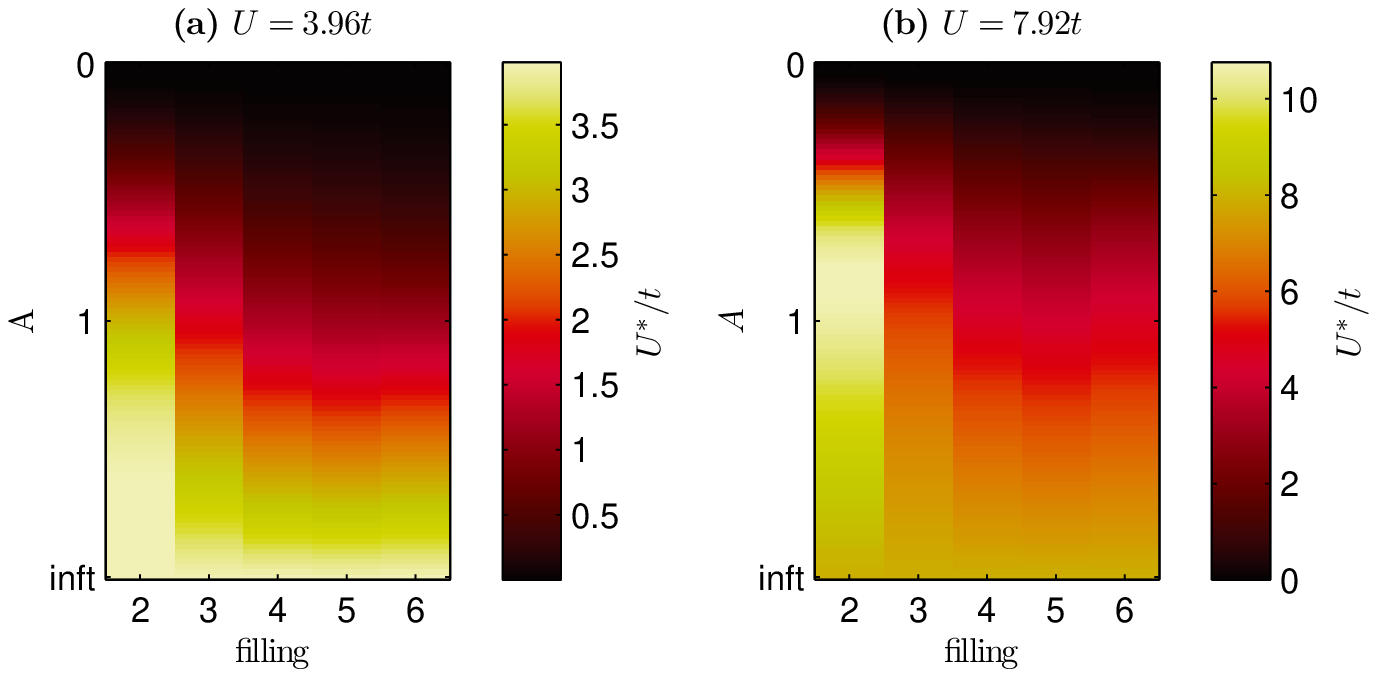}
}
\end{center}
\caption{(Color online) Effective local Coulomb interaction $U^*/t$ color coded for (a) benzene with $U=3.96/t$ and (b) benzene with $U=7.92/t$ , various screenings of $V_{ij}(\varepsilon)$ and all fillings. Due to particle hole symmetry of the model only $N\leq 6$ is shown.}
\label{fig:Ustarvsfilling}
\end{figure}
\begin{table}[htb]
\caption{First three rows: Coulomb matrix elements obtained with cRPA (graphene and silicene) and from \cite{bursill_optimal_1998} for benzene ($t_\text{graphene}=2.80 \eV$, $t_\text{silicene}=1.14 \eV$, $t_\text{benzene}=2.54 \eV$). Last three rows: Effective local Coulomb matrix elements for half filling with and without the approximation that electrons are only displaced to nearest neighbors and factor by which the local Coulomb interaction is decreased.}
\begin{ruledtabular}
\begin{tabular}[b]{lccc}
 & Graphene & Silicene &Benzene \\ \colrule
$U/t$      & $3.63$ & $ 4.19$ & $3.96$\\
$(V_{01}$,$V_{02})/t$ & $2.03$,$\ 1.45$ & $2.31$,$\ 1.72$ & $ 2.83$,$\ 2.01$\\
$(V_{03}$,$V_{04})/t$ & $1.32$,$\ 1.14$ & $1.55$,$\ 1.42$ & $1.80$,$\ -$ \\ \colrule
$U^*/t$       & $ 1.6\pm0.2 $ & $2.0 \pm 0.3 $ & $ 1.2 $\\
$(U-V_{01})/t$ & $1.6$ 	    & $1.9 $ & $1.1$\\
$U^*/U$	   & $0.45\pm 0.05$		 & $0.46 \pm 0.05 $   & $0.3$
\end{tabular}
\label{tab:results}
\end{ruledtabular}
\end{table}

It is interesting to see how the renormalization of the local Coulomb interaction depends on the filling of the system. Therefore, we study the model of benzene at arbitrary number of electrons $N$ by means of ED. The initial Coulomb matrix elements entering \mbox{Eq. (\ref{eq:origHam})} are assumed to be doping independent. The results for the filling dependent $U^*/t$ in benzene for different strengths of the nonlocal Coulomb interaction $V_{ij}(\varepsilon)$ from Eq. (\ref{eq:ohno}) are shown in Fig. \ref{fig:Ustarvsfilling}. Clearly doping in the range of $4\le N \le 8$ has only little effect on $U^*$. This doping range corresponds to changing the number of electrons on the order of $\pm 1/3$ per atom and thus covers fully the range \ms{of} dopings which can be achieved in graphene by means of gate voltages or adsorbates.

Strong differences to the half-filled case arise however for extreme doping (N=2,10), i.e. close to the nearly empty / almost completely filled case. The reduction of the effective local interaction $U^*$ is considerably weaker (Fig. \ref{fig:Ustarvsfilling}a)). For a stronger initial on-site interaction ($U = 7.92t$) $U^*$ even exceeds the initial on-site interaction by a factor of up to $U^*/U\approx1.3$ (Fig. \ref{fig:Ustarvsfilling}b)).
The physical origin of the behavior is illustrated in  Fig. \ref{fig:physical}b). In a dilute system, two electronic wave packages can minimize their Coulomb energy by simply avoiding each other in real space while staying delocalized over many lattice spacings at the same time. For such delocalized wave packages the effect of on-site and e.g. nearest neighbor Coulomb interactions becomes very similar and the on-site interaction $U^*$ is increased by $V$.

This effect, can be generally expected in nearly empty and almost filled systems: Fig. \ref{fig:corrFunc}b) shows the $U^*$ derivatives of charge correlation functions in a $5\times 5$ supercell of a honeycomb lattice occupied by $N=2$ electrons in total. Most importantly, $\partial_{U^*}\expec{n_{0\msu}n_{j\msd}}^*$ shows pronounced differences to the half filled case. In addition to a suppression of double occupancy by increased $U^*$ (i.e. $\partial_{U^*}\expec{n_{0\msu}n_{0\msd}}^*<0$ as in the half filled case) $\partial_{U^*}\expec{n_{0\msu}n_{j\msd}}^*$ is negative in the vicinity of $j=0$, too. Increasing local interactions with an electron at site $j=0$ expel other electrons also from its vicinity. This corresponds to the process depicted in Fig. \ref{fig:physical}b) and leads to effective on-site interactions being \textit{increased} by non-local Coulomb terms. This can be understood in terms of Wigner crystallization \cite{wigner_interaction_1934}. In the full model the Coulomb energy wins over the kinetic energy for low electron/hole densities. Thus the carriers tend to localize. To approach a Wigner crystal also in the auxiliary model, the effective local interaction is increased such that interaction energy dominates over kinetic energy.

Finally, the question arises how accurate the effective model reflects the physical properties of the original model. The phase diagram of the extended Hubbard model on the honeycomb lattice includes an antiferromagnetic (AF), a semimetal (SM) and a charge density wave (CDW) phase \cite{herbut_interactions_2006}, while the Hubbard model with strictly local interactions only features the first two phases. Similarly, if the system is in a quantum Hall regime, i.e. presence of strong magnetic fields, there are some many-body phenomena like the formation of stripes where the long-range tails of the Coulomb interaction are crucially important. In situations with such charge inhomogeneities the auxiliary model can likely fail to provide a physically correct description of the original system. If the parameters of the extended model are, however, clearly inside the AF or the SM phase, the effective model will likely approximate the physical properties of the original model quite well. 

\begin{figure}[t]
\begin{center}
\mbox{
\psfrag{inf}{$\infty$}
\psfrag{A}{$\varepsilon$}
\includegraphics[width=1\columnwidth]{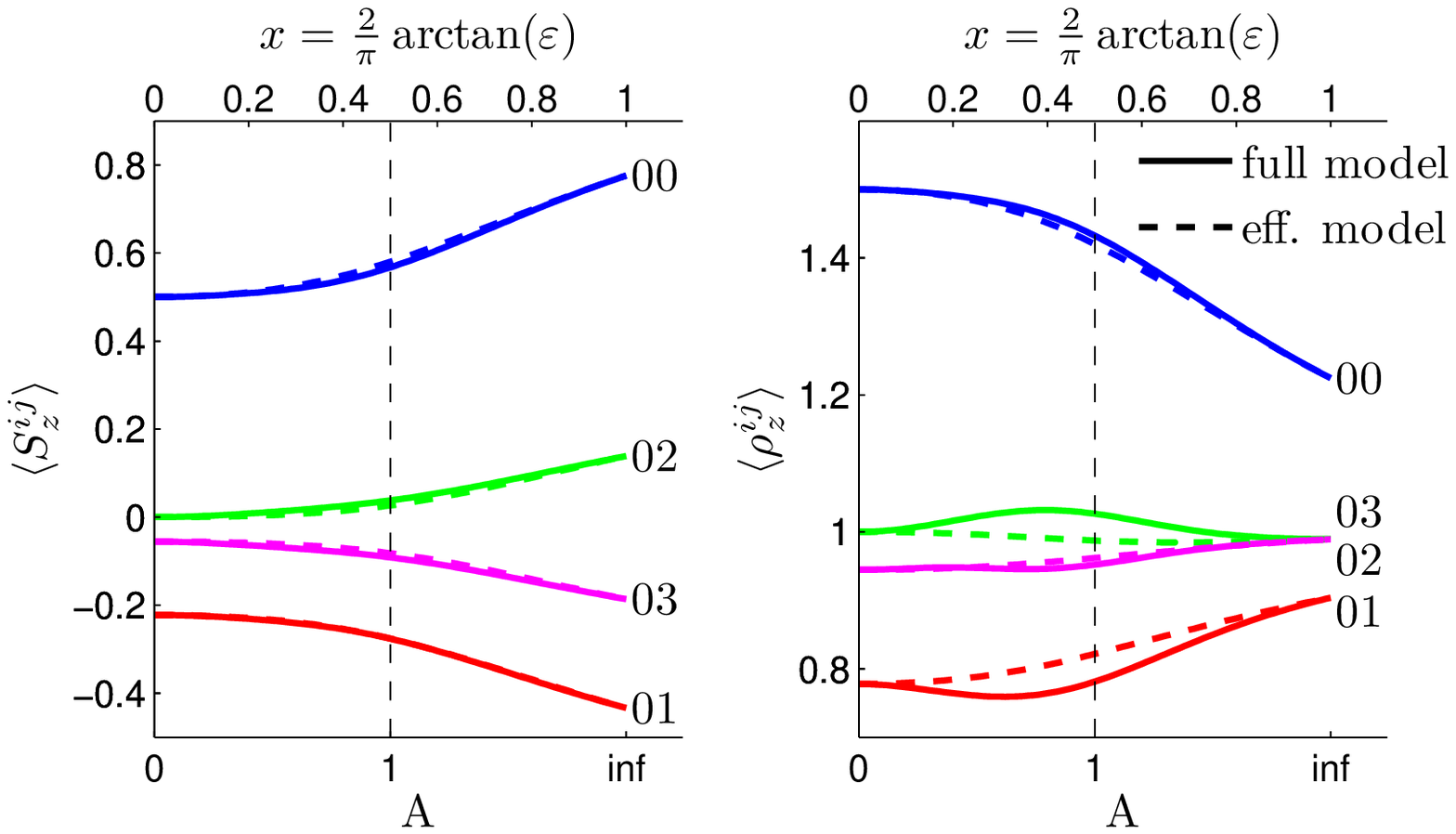}
}
\end{center}
\caption{(Color online) Correlation as functions of the screening for the extended Hubbard model (continuous lines) and the effective Hubbard model (broken lines) for benzene. The left panel shows spin correlation $\langle S^{ij}_z \rangle$ and the right panel shows density correlation $\langle \rho^{ij}_z \rangle$. $\langle S^{01}_z\rangle$ is virtually the same for the effective and original model. The parameters for the original model are $U = 10.06 \eV$, $t_\text{orig}=2.539 \eV$ and $V(\varepsilon)$ is calculated by Eq. (\ref{eq:ohno}). $U^*(\varepsilon)$ is calculated by (\ref{eq:Ustar}), while $t_\text{eff}=t_\text{orig}$.}
\label{fig:corr}
\end{figure}

We illustrate this expectation with the example of modified benzene. In this model, the nonlocal Coulomb interaction $V_{ij}$ are calculated with the Ohno interpolation (\ref{eq:ohno}). A comparison of the spin $\langle S^{ij}_z \rangle = \langle(n_{i\msu}-n_{i\msd})(n_{j\msu}-n_{j\msd}) \rangle$ and the density correlation functions $\langle \rho^{ij} \rangle = \langle(n_{i\msu}+n_{i\msd})(n_{j\msu}+n_{j\msd}) \rangle$ for the extended and the auxiliary local Hubbard model are shown in Fig. \ref{fig:corr}. The correlation functions have been calculated by exact diagonalization for, both, the original and the effective model. For $\varepsilon=0$ and $\varepsilon\rightarrow\infty$ (non-interacting \ms{and local} limit\ms{, respectively}) the correlation functions of the effective and original model coincide as they should. CDW physics would manifest in $\langle \rho^{ij} \rangle$ and here we find indeed some differences of $\langle \rho^{ij} \rangle$ for the effective and the auxiliary model for intermediate screening ($\epsilon \sim 1$). However, nearly no deviation of $\langle S^{ij}_z \rangle$ between the extended and effective model is found. This behavior is found for all fillings and also different initial local interactions U \cite{suppl}. We thus expect that transitions into phases like an AF insulator (or a Mott insulator) will be very well described by the effective model.

In conclusion, a systematic map from lattice models with nonlocal Coulomb interactions to effective Hubbard models with strictly local Coulomb interactions $U^*$ is derived. The physical properties of the effective model reflect the original system nicely, especially regarding spin related properties. We find that the nonlocal Coulomb interactions can significantly renormalize the effective on-site interaction $U^*$ as compared to the original local $U$. In the cases of graphene and silicene our calculations yield $U^*/U<0.5$ for half filling. Thus, the nonlocal Coulomb interactions stabilize the Dirac semimetallic phases in these materials against transitions to a gapped spin liquid or an antiferromagnetic insulator. In defective graphene or at edges local Coulomb interactions can lead to the formation of magnetic moments \cite{Misha_book,lopez-sancho_magnetic_2009,yazyev_emergence_2010}. When describing these situations in terms of the Hubbard model, the value of $U^* = 1.6t$ obtained here should be used. Whether or not a Hubbard model is generally appropriate to describe the physical properties of graphene is still a matter of debate and depends on the observable of interest. Our results suggest that a Hubbard model should be useful to judge the occurrence of edge magnetism and of AF insulator phases. Furthermore, our work indicates that nonlocal Coulomb interactions will, in general, significantly weaken local correlation effects in $sp$-electron materials in a wide doping range. Additionally we have shown that for extreme low carrier densities (in the vicinity of the Wigner crystal instability) non local interactions can increase the effective local interaction. Such systems should be realizable e.g. in any weakly doped semiconductor. It is interesting to see how the renormalization of effective on-site interactions generalizes to heterostructures with modified bands and additional van Hove singularities like in twisted bilayer graphene, gated (gapped) bilayer or to quantum Hall systems depending on Landau level filling factors.

\textit{Acknowledgements.} The authors thank R. Scalletar for help with the \textsc{quest} code, F. Assaad for providing his DQMC code and D. Mourad and F. Jahnke for helpful discussions. Financial support from DFG via SPP 1459 and FOR 1346 are acknowledged. MIK acknowledges a support from FOM (Netherlands).

\bibliographystyle{apsrev4-1}
\bibliography{BibliogrGrafeno}

\end{document}


\title{Supplemental Material: Optimal Hubbard models for materials with nonlocal Coulomb interactions: graphene, silicene and benzene}
\author{M. Sch\"uler}
\author{M. R\"osner}
\author{T. O. Wehling}
\affiliation{Institut f{\"u}r Theoretische Physik, Universit{\"a}t Bremen, Otto-Hahn-Allee 1, 28359 Bremen, Germany}
\affiliation{Bremen Center for Computational Materials Science, Universit{\"a}t Bremen, Am Fallturm 1a, 28359 Bremen, Germany}
\author{A. I. Lichtenstein}
\affiliation{Institut f{\"u}r Theoretische Physik,
Universit{\"a}t Hamburg, Jungiusstra{\ss}e 9, D-20355 Hamburg,
Germany}
\author{M. I. Katsnelson}
\affiliation{Radboud University of Nijmegen, Institute for
Molecules and Materials, Heijendaalseweg 135, 6525 AJ Nijmegen,
The Netherlands}

\pacs{72.80.Rj; 73.20.Hb; 73.61.Wp}
\date{\today}

\begin{abstract}
In this supporting material we provide additional detail regarding technical details of our calculations. The convergence with respect to real space cut-offs is discussed. In addition a filling dependent comparison of spin and density correlation functions derived from the auxiliary and original model is given.
\end{abstract}

\maketitle

\section{Real space convergence of the effective interaction}
\begin{figure*}[t]
\begin{center}
\includegraphics[width=0.49\linewidth]{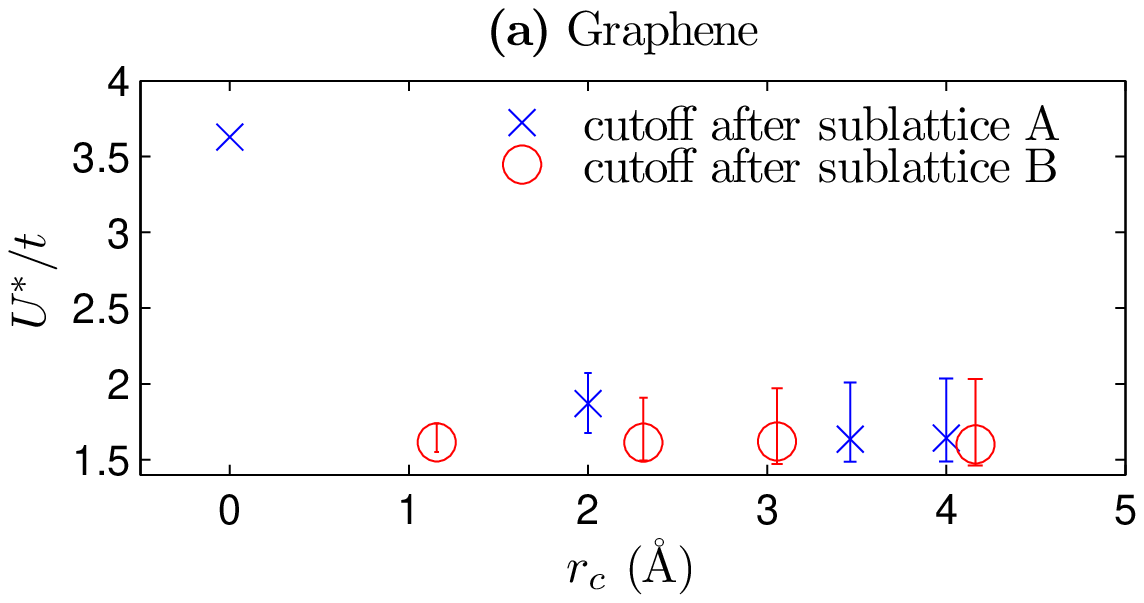}
\vspace{0pt}
\includegraphics[width=0.49\linewidth]{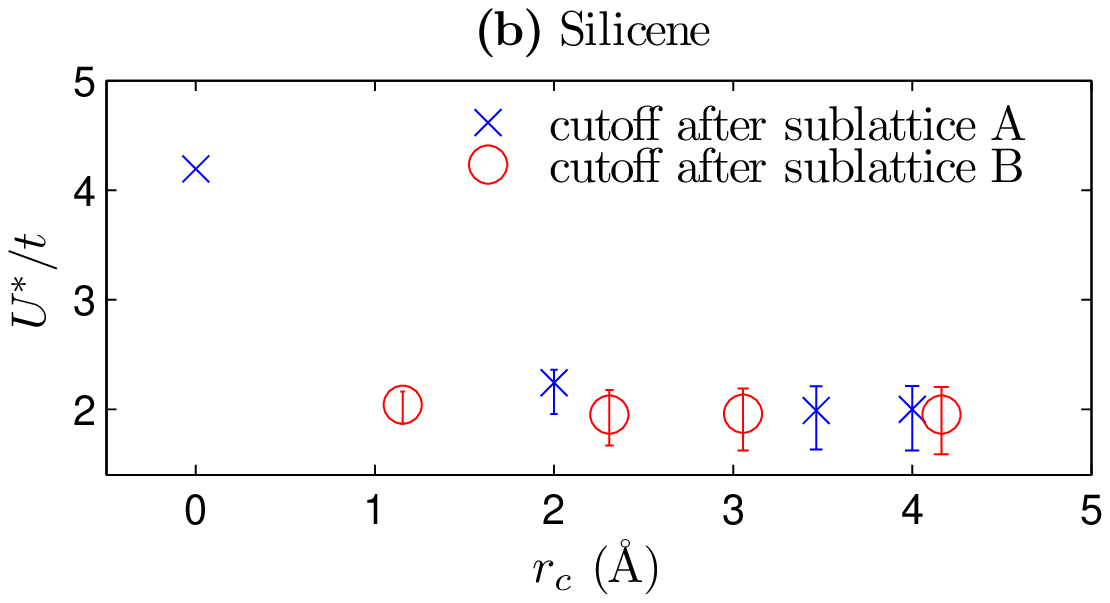}
\caption{(Color online) Effective Coulomb interaction $U^*$ for graphene and silicene against the position of the cutoff used to evaluate Eq. (5).}
\end{center}
\label{fig:UstarVsR}
\end{figure*}

For translational invariant systems the equation for the effective interaction (Eq. (5) in the main paper) includes a sum of derivatives of the correlation functions times the non-local Coulomb interaction over all neighbors $j\neq0$. The spacial convergence depends on the $r$ dependence of the correlation function and that of the nonlocal Coulomb interaction $V_{0j}$. The Coulomb interaction decays asymptotically like $1/r$. The derivatives fall off rather quickly and their sign changes with the sublattice. Altogether a quick convergence can be expected: Fig. \ref{fig:UstarVsR} shows the results of the evaluation of Eq. (5) for different cutoff radii $r_c$ for graphene and silicene respectively. $r_c$ is the distance between the sites with indexes $i=0$ and $j$. The error bars are determined by the statistical errors of the DQMC calculations. As could be already expected from the $\partial_{U^*}\expec{n_{0\msu}n_{j\sigma'}}^*$-terms depicted in Fig. 2, the nearest neighbor Coulomb interaction has the strongest impact on the renormalization of the on-site term $U^*$. Thus, the approximation of neglecting all non nearest neighbors terms is quite reasonable, here. Contributions from the same sublattice as the site $i=0$ tend to increase $U^*$ and those from the other sublattice decrease $U^*$. It is obviously sufficient to introduce the cutoff after the fourth nearest neighbors.

\section{Reliability of the method in dependence of the filling for benzene}
\begin{figure*}[t]
\begin{center}
\psfrag{inft}{$\infty$}
\includegraphics[width=0.45\linewidth]{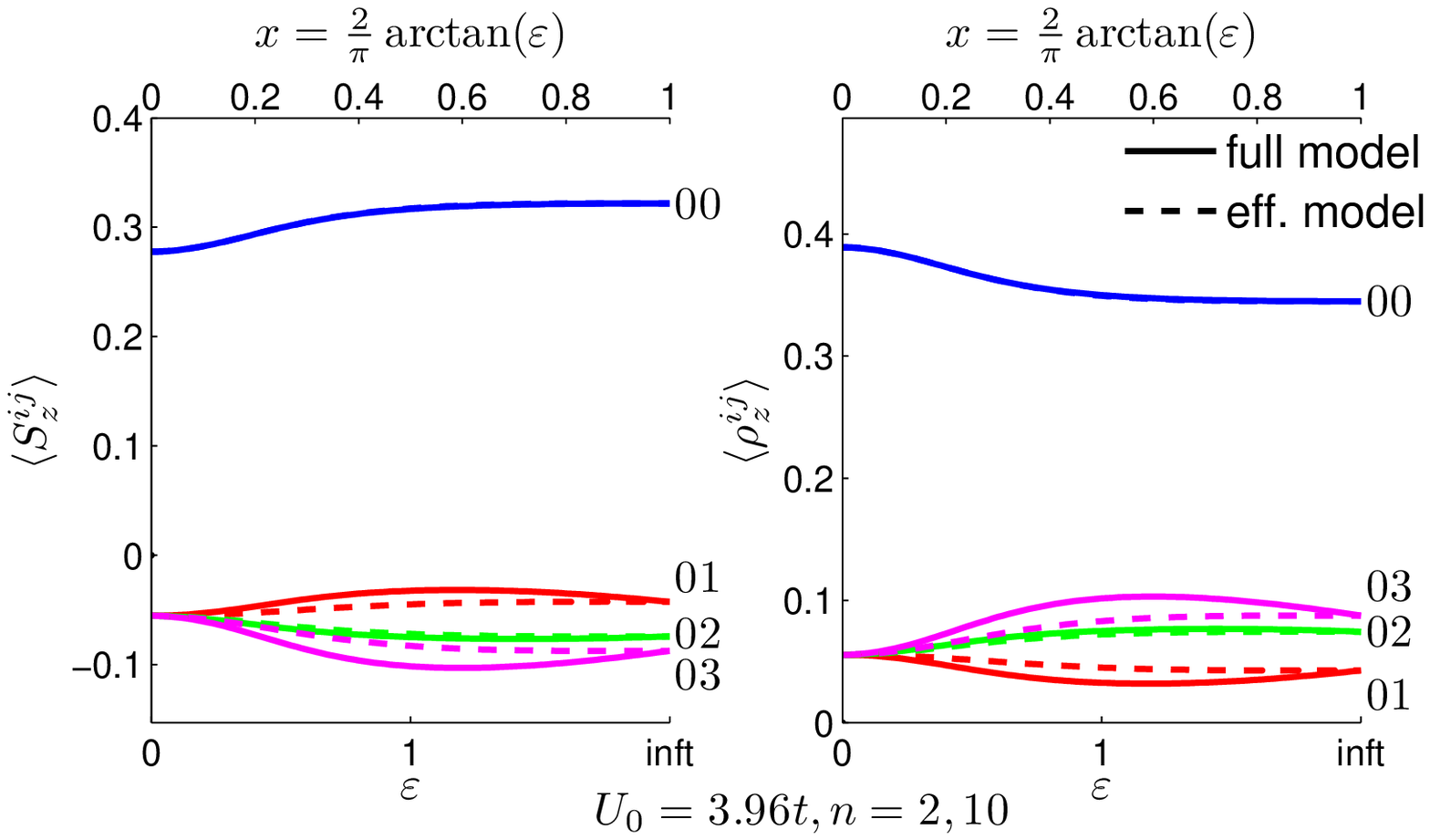}
\psfrag{inft}{$\infty$}
\includegraphics[width=0.45\linewidth]{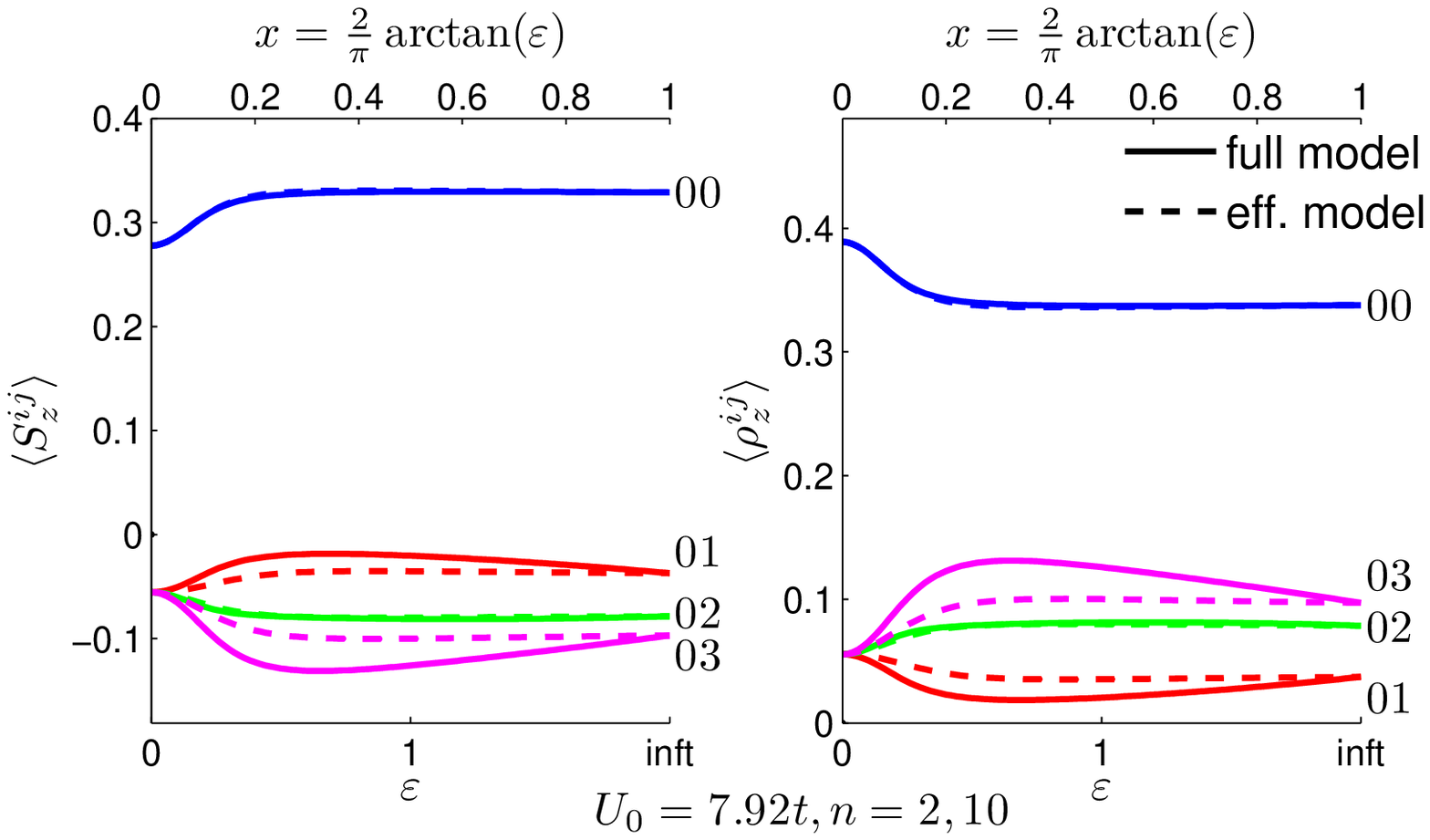}
\hspace{8pt}%
\psfrag{inft}{$\infty$}
\includegraphics[width=0.45\linewidth]{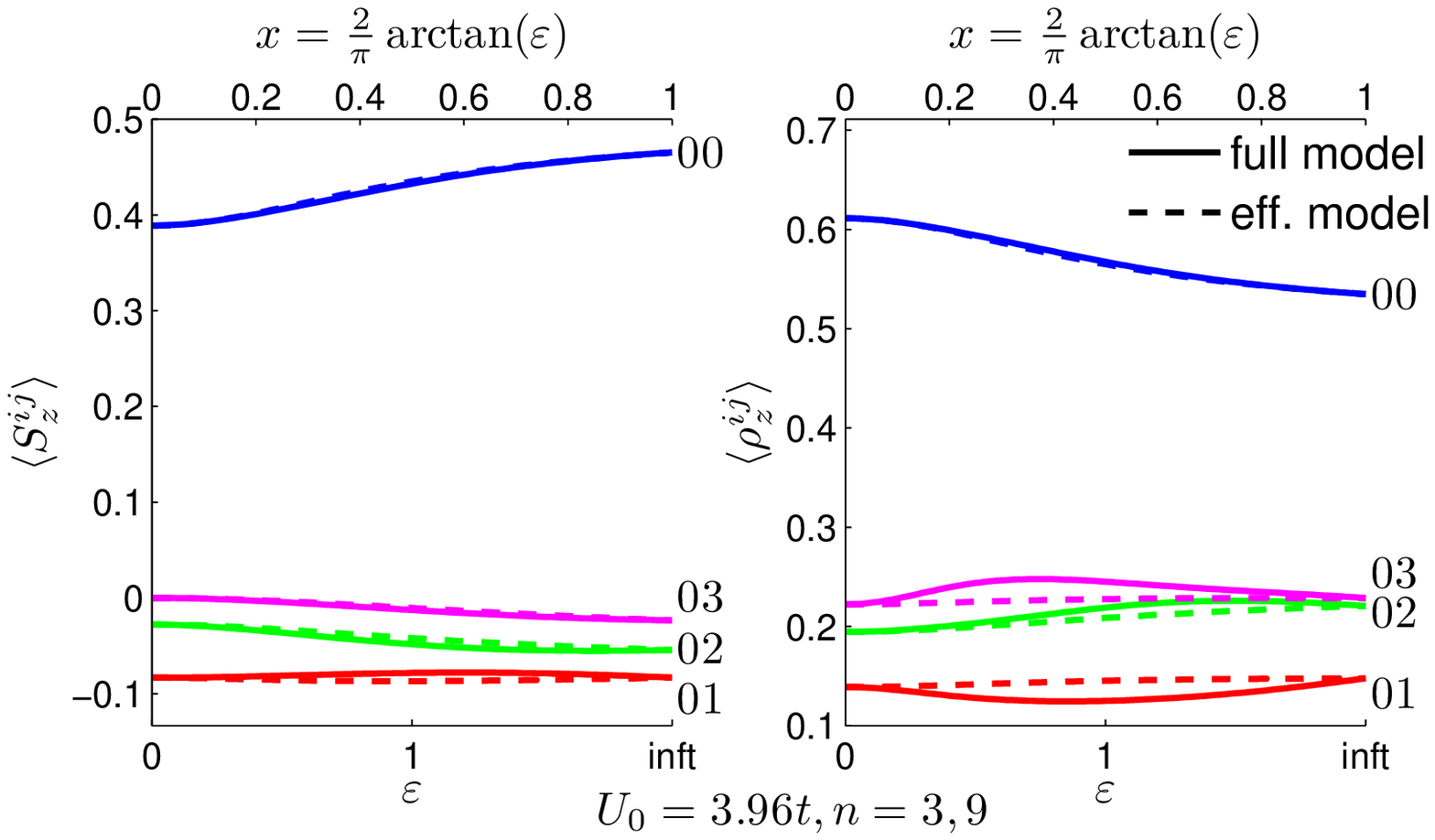}
\psfrag{inft}{$\infty$}
\includegraphics[width=0.45\linewidth]{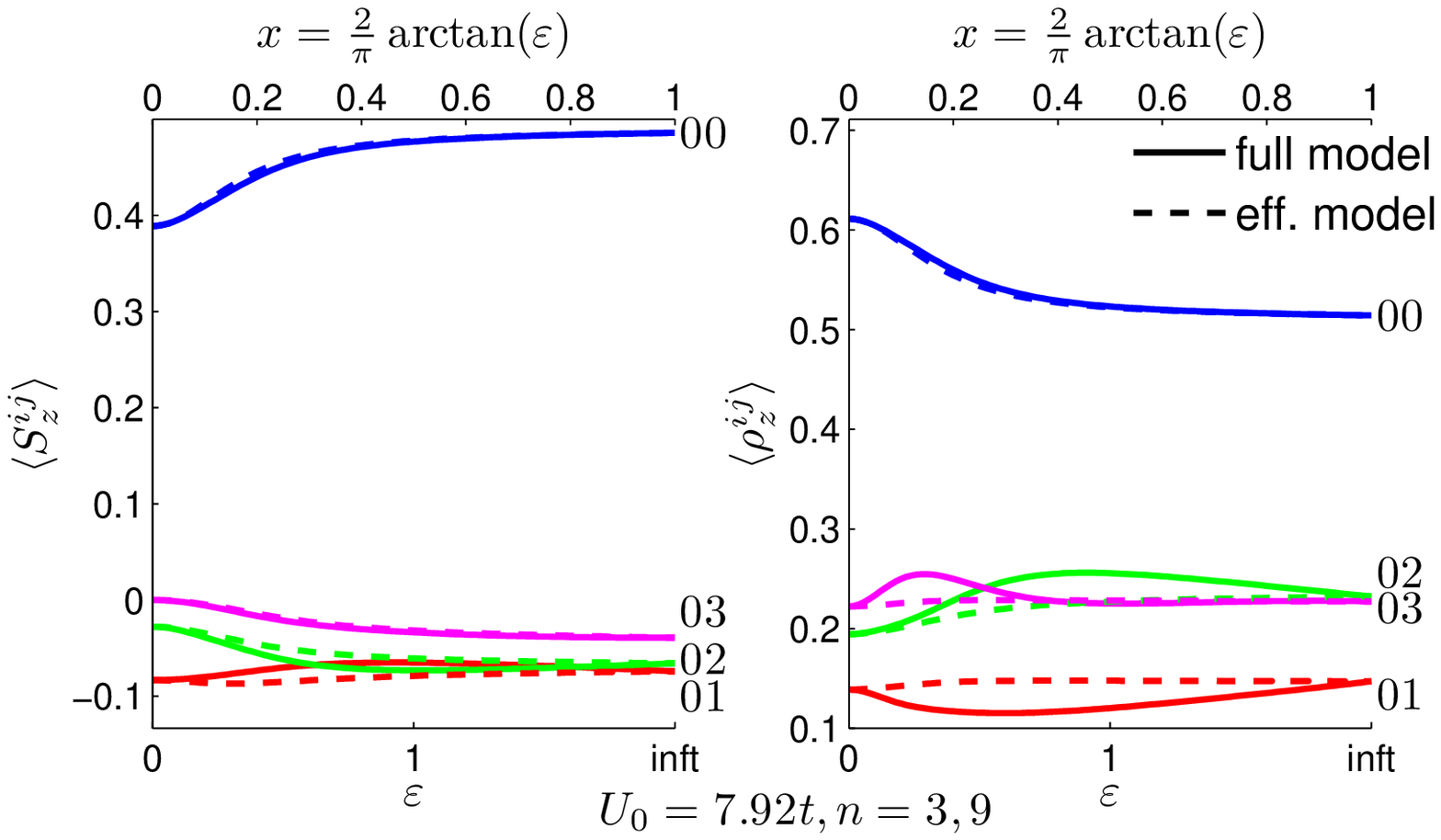}
\hspace{8pt}%
\psfrag{inft}{$\infty$}
\includegraphics[width=0.45\linewidth]{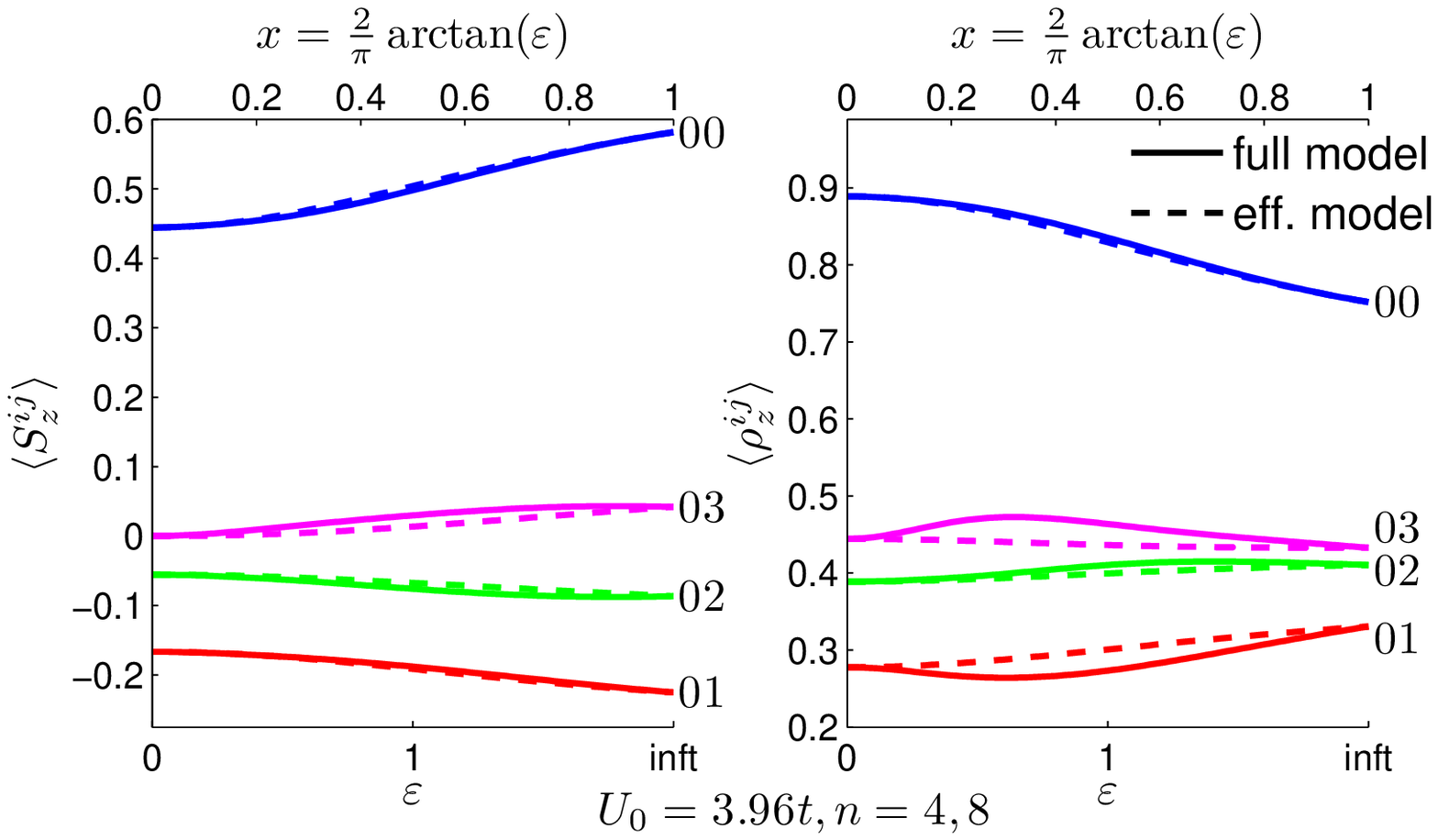}
\psfrag{inft}{$\infty$}
\includegraphics[width=0.45\linewidth]{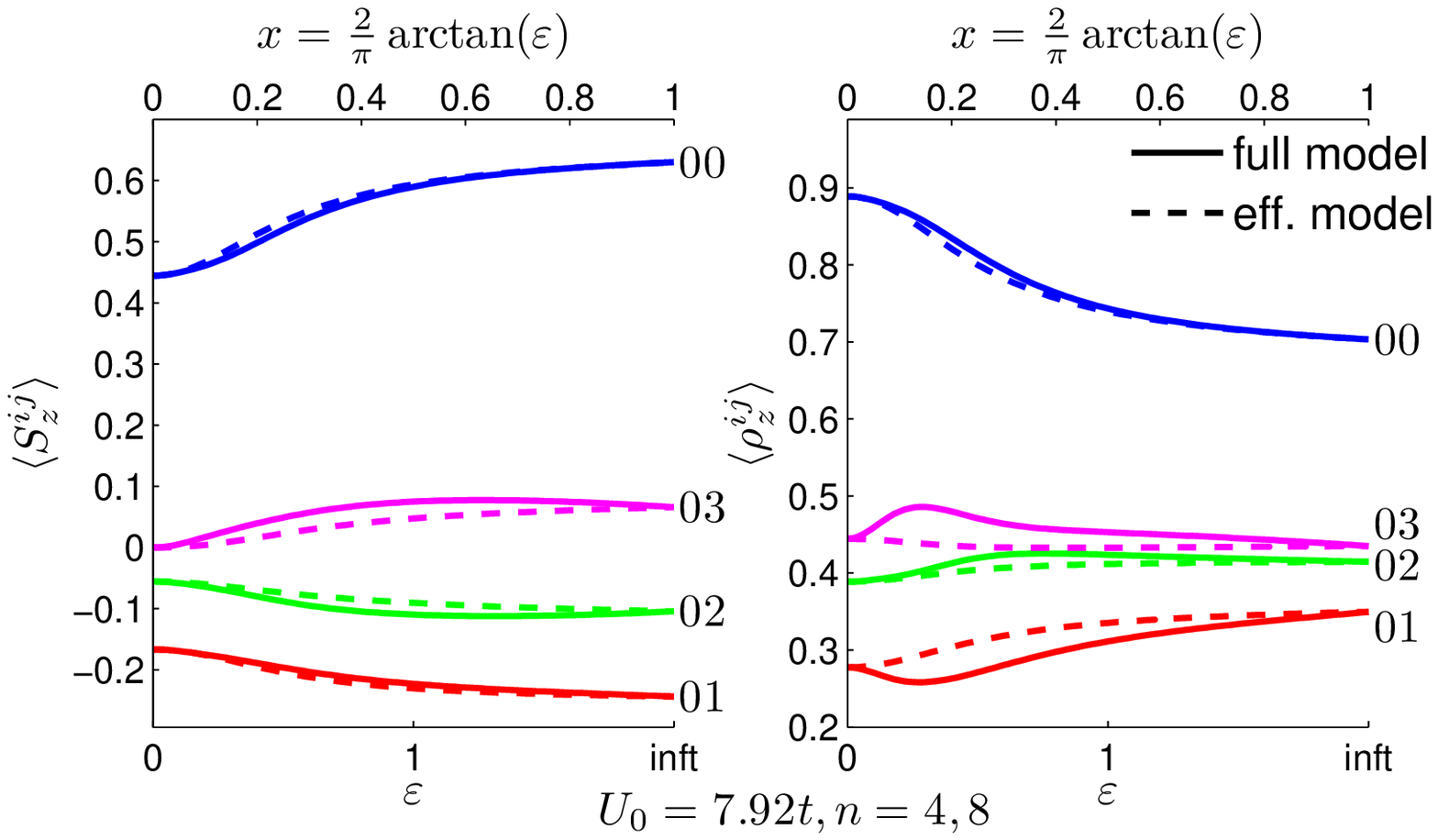}
\hspace{8pt}%
\psfrag{inft}{$\infty$}
\includegraphics[width=0.45\linewidth]{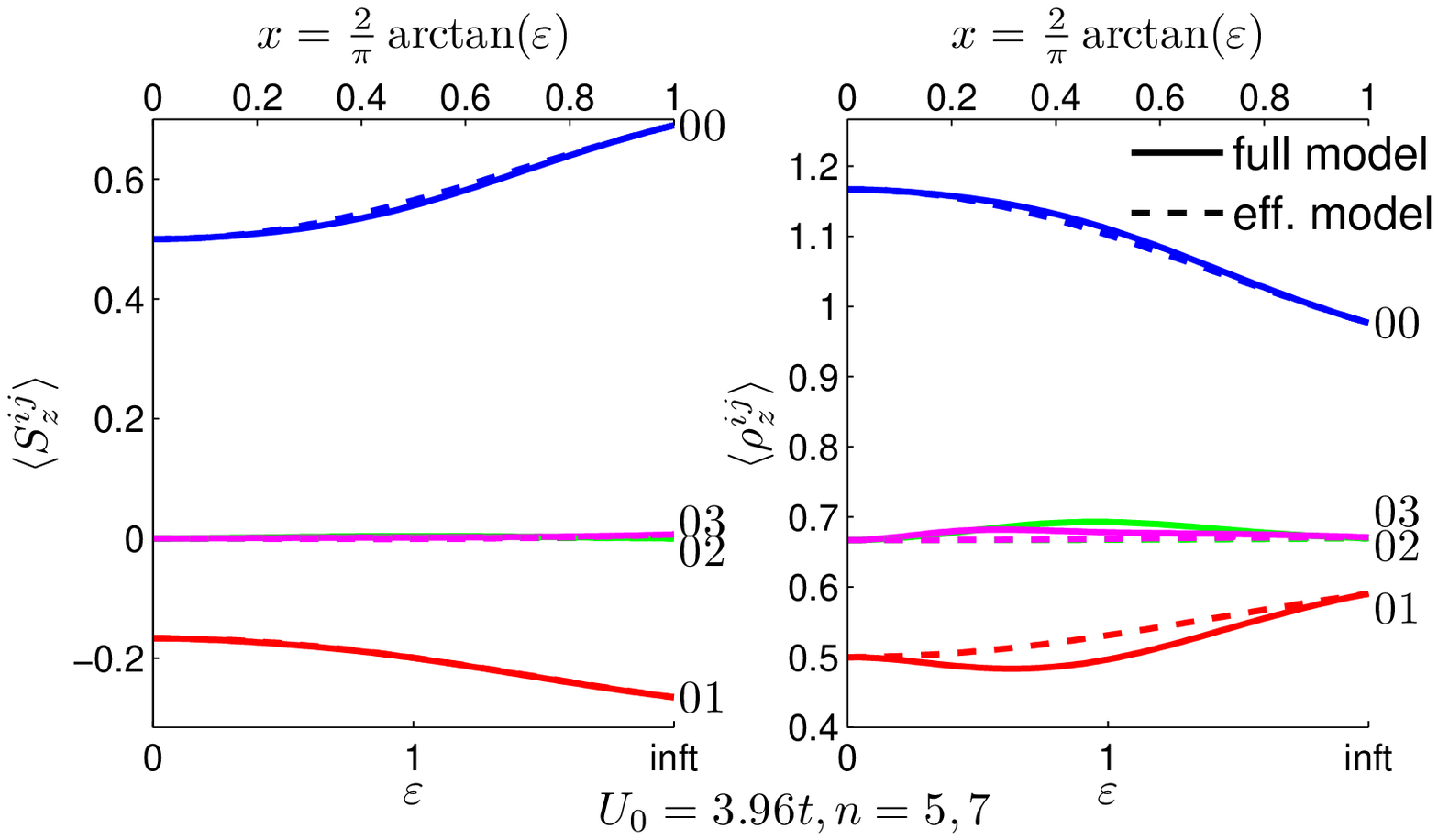}
\psfrag{inft}{$\infty$}
\includegraphics[width=0.45\linewidth]{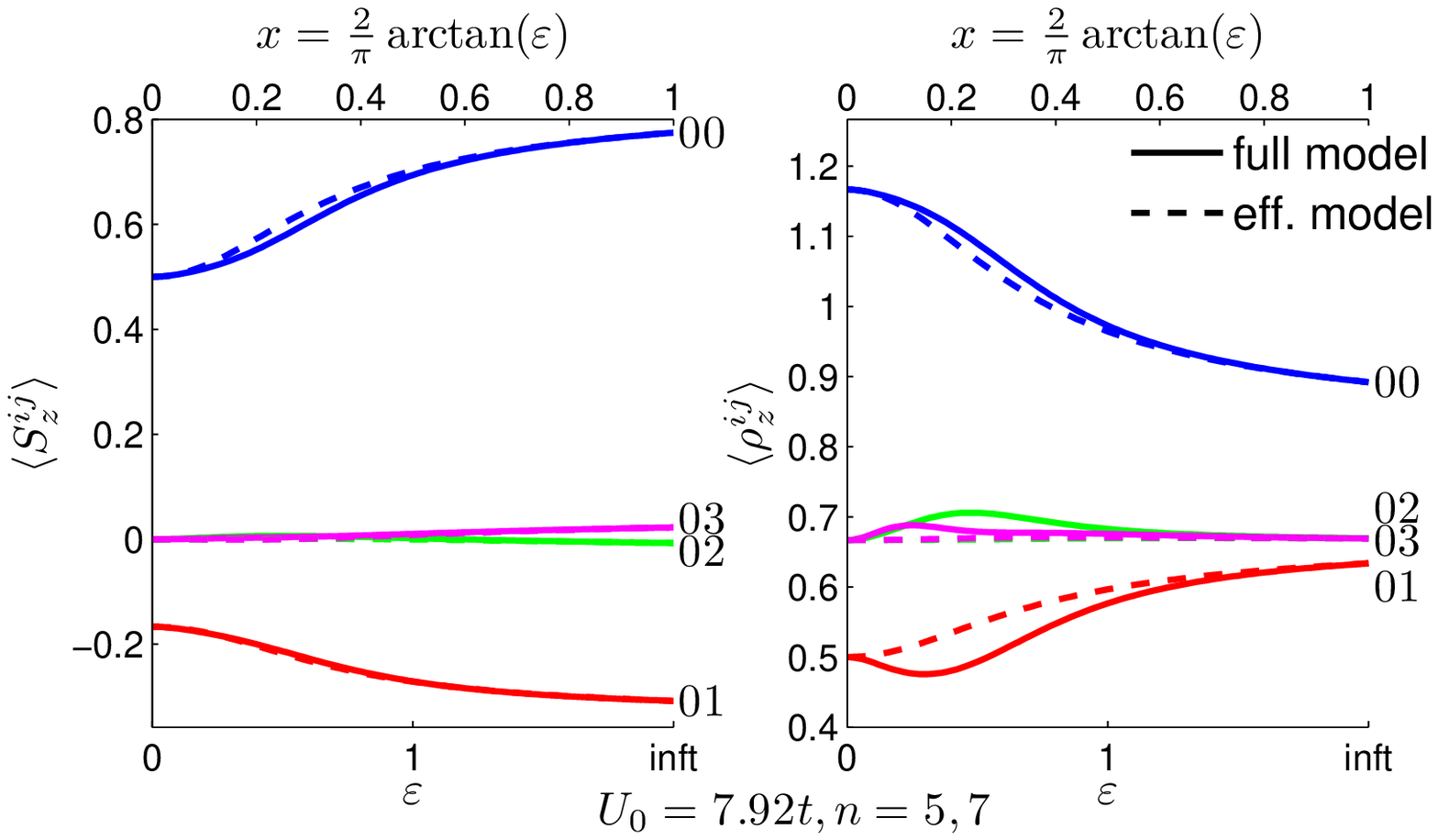}
\hspace{8pt}%
\psfrag{inft}{$\infty$}
\includegraphics[width=0.45\linewidth]{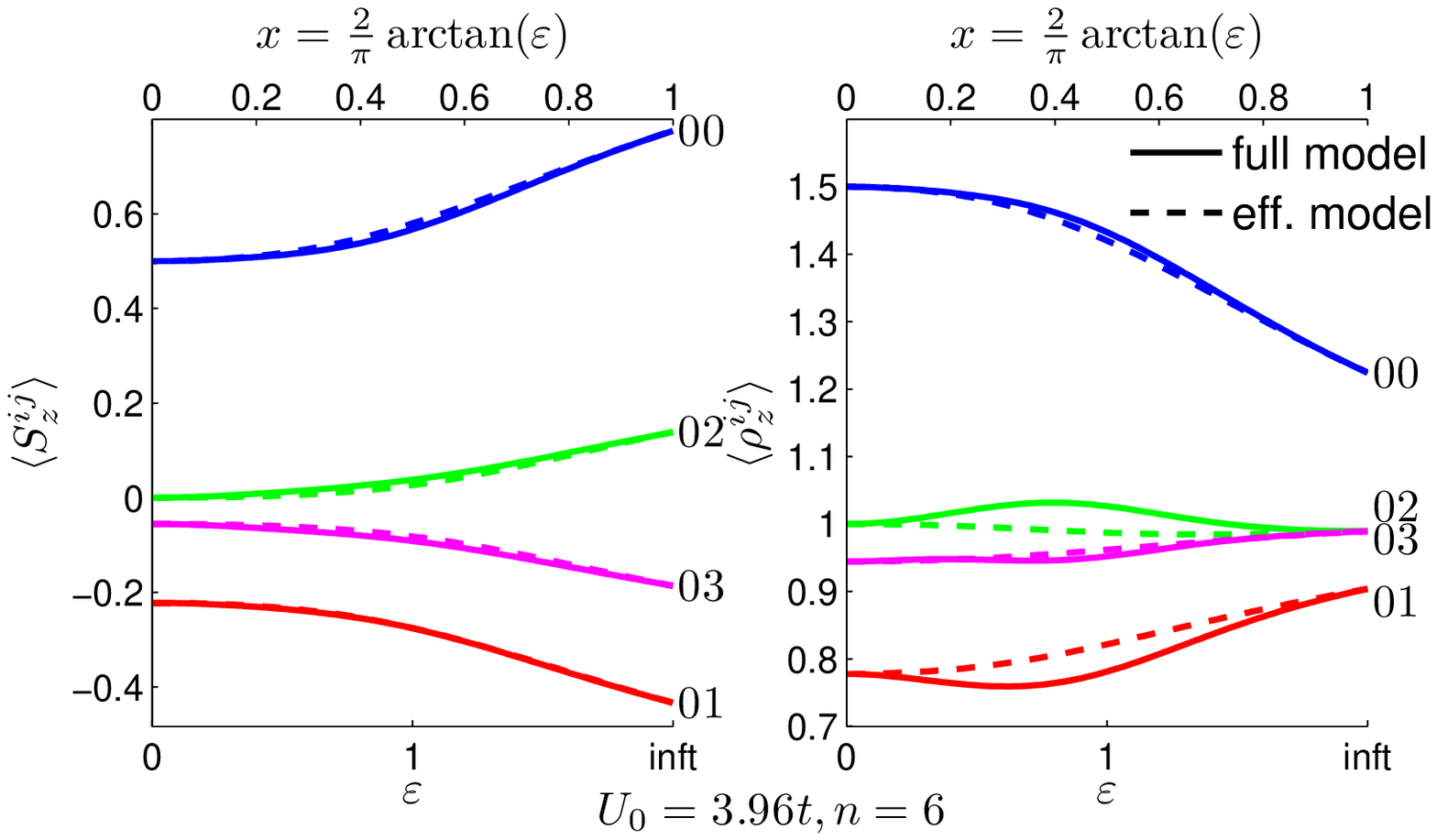}
\psfrag{inft}{$\infty$}
\includegraphics[width=0.45\linewidth]{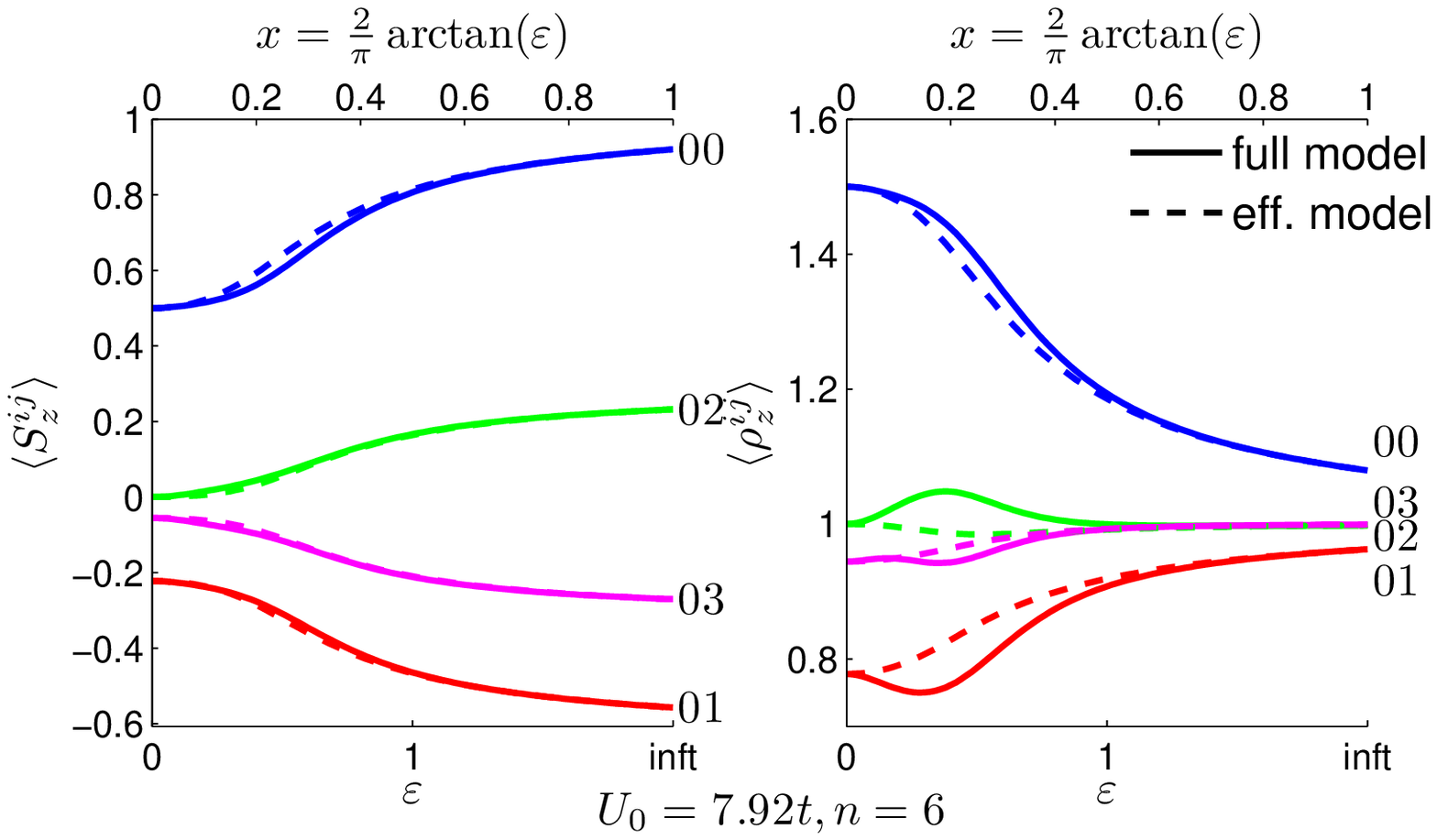}
\caption{Comparison of the spin and density correlation function for modified benzene. The left panels show results for $U=3.96t$, the right panels show results for $U=7.92t$. From top to bottom, the panels show results for different fillings.}
\label{fig:comp}
\end{center}
\end{figure*}

This section provides a comparison of the spin and density correlation functions for the auxiliary and original model, to judge the reliability of our method for benzene. Fig. \ref{fig:comp} shows the spin correlation function and density correlation function for all fillings and two different values of the original Coulomb interaction $U$ for the effective and original model. The fact, that the spin correlation is reflected better than the density correlation is valid for all fillings and both $U$ values. However, both the spin and the density correlation function of the effective system show only small deviations from the original ones. Thus, the effective model reflects the physical properties of the original model nicely for all fillings. 

